\newif\ifREFEREE
 \REFEREEtrue
 \REFEREEfalse

\ifREFEREE
	\documentstyle[epsf,referee]{mn}
\else
	\documentstyle[epsf]{mn}
\fi

\newif\ifAMStwofonts
\AMStwofontsfalse

\def\h{\rmn{h}}

\def\dsy{\displaystyle}

\def\beq{\begin{equation}}
\def\eeq{\end{equation}}
\def\beqn{\begin{eqnarray}}
\def\eeqn{\end{eqnarray}}

\def\sech{\mathop{\rm sech}\nolimits}

\def\vc{v_\rmn{c}}
\def\D{{\rmn{d}}}
\renewcommand{\d}{\rmn{d}}
\def\p{\upartial}

\def\masyr{\mbox{$\,{\rm mas}\,{\rm yr}^{-1}$}}
\def\kms{\mbox{$\,{\rm km}\,{\rm s}^{-1}$}}
\def\kpc{\mbox{$\,{\rm kpc}$}}
\def\parc{\mbox{$\,{\rm pc}$}}
\def\Msun{\mbox{$\,{\rm M}_\odot$}}
\let\msun\Msun
\def\HI{\hbox{\rm H\thinspace\sc i}}
\def\HII{\hbox{\rm H\thinspace\sc ii}}
\def\SgrA{\hbox{\rm Sgr\thinspace A$^\star$}}

\def\etal{\mbox{\it et~al.}}

\def\refer#1#2#3{#1, #2, #3}

\def\aaa#1#2{\refer{A\&A}{#1}{#2}}
\def\aas#1#2{\refer{A\&AS}{#1}{#2}}

\def\aj#1#2{\refer{AJ}{#1}{#2}}

\def\apj#1#2{\refer{ApJ}{#1}{#2}}
\def\apjs#1#2{\refer{ApJS}{#1}{#2}}

\def\mn#1#2{\refer{MNRAS}{#1}{#2}}

\ifoldfss
  \newcommand{\rmn}[1] {{\rm #1}}

  \ifCUPmtlplainloaded \else
    \NewTextAlphabet{textbfit} {cmbxti10} {}
    \NewTextAlphabet{textbfss} {cmssbx10} {}
    \NewMathAlphabet{mathbfit} {cmbxti10} {} % for math mode
    \NewMathAlphabet{mathbfss} {cmssbx10} {} %  "   "    "
  \fi
  \ifAMStwofonts
    \ifCUPmtlplainloaded \else
      \NewSymbolFont{upmath} {eurm10}
      \NewSymbolFont{AMSa} {msam10}
      \NewMathSymbol{\upi}     {0}{upmath}{19}
      \NewMathSymbol{\umu}     {0}{upmath}{16}
      \NewMathSymbol{\upartial}{0}{upmath}{40}
      \NewMathSymbol{\leqslant}{3}{AMSa}{36}
      \NewMathSymbol{\geqslant}{3}{AMSa}{3E}

       \let\le=\leqslant
       \let\ge=\geqslant
    \fi
  \fi
\fi % End of OFSS

\ifnfssone
  \newmathalphabet{\mathit}
  \addtoversion{normal}{\mathit}{cmr}{m}{it}
  \addtoversion{bold}{\mathit}{cmr}{bx}{it}
  \newcommand{\rmn}[1] {\mathrm{#1}}

  \newmathalphabet{\mathbfit} % math mode version of \textbfit{..}
  \addtoversion{normal}{\mathbfit}{cmr}{bx}{it}
  \addtoversion{bold}{\mathbfit}{cmr}{bx}{it}
  \newmathalphabet{\mathbfss} % math mode version of \textbfss{..}
  \addtoversion{normal}{\mathbfss}{cmss}{bx}{n}
  \addtoversion{bold}{\mathbfss}{cmss}{bx}{n}
  \ifAMStwofonts
    \ifCUPmtlplainloaded \else
      \UseAMStwoboldmath
      \makeatletter
      \new@mathgroup\upmath@group
      \define@mathgroup\mv@normal\upmath@group{eur}{m}{n}
      \define@mathgroup\mv@bold\upmath@group{eur}{b}{n}
      \edef\UPM{\hexnumber\upmath@group}
      \new@mathgroup\amsa@group
      \define@mathgroup\mv@normal\amsa@group{msa}{m}{n}
      \define@mathgroup\mv@bold\amsa@group{msa}{m}{n}
      \edef\AMSa{\hexnumber\amsa@group}
      \makeatother
      \mathchardef\upi="0\UPM19
      \mathchardef\umu="0\UPM16
      \mathchardef\upartial="0\UPM40
      \mathchardef\leqslant="3\AMSa36
      \mathchardef\geqslant="3\AMSa3E

       \let\le=\leqslant
       \let\ge=\geqslant
    \fi
  \fi
\fi % End of NFSS release 1

\ifnfsstwo
  \newcommand{\rmn}[1] {\mathrm{#1}}

  \DeclareMathAlphabet{\mathbfit}{OT1}{cmr}{bx}{it}
  \SetMathAlphabet\mathbfit{bold}{OT1}{cmr}{bx}{it}
  \DeclareMathAlphabet{\mathbfss}{OT1}{cmss}{bx}{n}
  \SetMathAlphabet\mathbfss{bold}{OT1}{cmss}{bx}{n}
  \ifAMStwofonts
    \ifCUPmtlplainloaded \else
      \DeclareSymbolFont{UPM}{U}{eur}{m}{n}
      \SetSymbolFont{UPM}{bold}{U}{eur}{b}{n}
      \DeclareSymbolFont{AMSa}{U}{msa}{m}{n}
      \DeclareMathSymbol{\upi}{0}{UPM}{"19}
      \DeclareMathSymbol{\umu}{0}{UPM}{"16}
      \DeclareMathSymbol{\upartial}{0}{UPM}{"40}
      \DeclareMathSymbol{\leqslant}{3}{AMSa}{"36}
      \DeclareMathSymbol{\geqslant}{3}{AMSa}{"3E}

       \let\le=\leqslant
       \let\ge=\geqslant
    \fi
  \fi
\fi % End of NFSS release 2

\ifCUPmtlplainloaded \else
  \ifAMStwofonts \else % If no AMS fonts
    \def\upi{\pi}
    \def\umu{\mu}
    \def\upartial{\partial}
  \fi
\fi

\begin{document}

\title{Mass models of the Milky Way}

\author[W.\ Dehnen and J.\ Binney]
{	Walter Dehnen and James Binney	\\
Theoretical Physics, 1 Keble Road, Oxford OX1 3NP}

%\date{Accepted  Received ; in original form}
\pubyear{1997}

\maketitle

\begin{abstract}
A parameterized model of the mass distribution within the Milky Way is fitted
to the available observational constraints. The most important single parameter 
is the ratio of the scale length $R_{\d,*}$ of the stellar disk to $R_0$. The 
disk and bulge dominate $\vc(R)$ at $R\la R_0$ only for $R_{\d,*}/R_0\la 0.3$.
Since the only knowledge we have of the halo derives from studies like the 
present one, we allow it to contribute to the density at all radii. When 
allowed this freedom, however, the halo causes changes in assumptions relating
to $R\ll R_0$ to affect profoundly the structure of the best-fitting model
at $R\gg R_0$. For example, changing the disk slightly from
an exponential surface-density profile significantly changes the form of
$\vc(R)$ at $R\gg R_0$, where the disk makes a negligible contribution to 
$\vc$. Moreover, minor changes in the constraints can cause the halo to develop
a deep hole at its centre that is not physically plausible. These problems call
into question the proposition that flat rotation curves arise because galaxies
have physically distinct halos rather than outwards-increasing mass-to-light
ratios.
\par
The mass distribution of the Galaxy and the relative importance of its various 
components will remain very uncertain until more observational data can be used
to constrain mass models. Data that constrain the Galactic force field at 
$z\ga R$ and at $R>R_0$ are especially important. 
\end{abstract}


\section{INTRODUCTION}
One of the fundamental tasks of Galactic astronomy is the determination of
the mass and luminosity distributions of the Milky Way. In the 1950s the
development of radio astronomy opened up the study of the Galaxy's
large-scale structure, and much of the understanding that was attained at
that time was summarized by Schmidt's (1956) mass model.  In the 1970s and
early 1980s our picture of the Milky Way changed in response both to studies
of external galaxies and to a growing awareness of the existence of ``dark
matter'' at large radii. These developments were reflected in the Bahcall \&
Soneira (1980), Caldwell \& Ostriker (1981) and Rohlfs \& Kreitschmann
(1988) Galaxy models. These models were based on two rather different
methodologies: whereas Bahcall \& Soneira concentrated on fitting the
distribution of {\it luminosity\/} within the Galaxy by fitting star counts,
Caldwell \& Ostriker and Rohlfs \& Kreitschmann concentrated on fitting
various measures of the Galactic gravitational force-field. Never the less,
all these models decompose the Galaxy into ``components'' that are
motivated by photometric studies of external galaxies, and incorporated a
range of dynamical constraints. It is our aim in this and subsequent papers
to update and extend these models.

The principal direction in which we wish to extend traditional galaxy models
is the incorporation of kinematic information that is capable of
constraining the degree of flattening of the mass distribution.  The
kinematic information that has traditionally been used to constrain galaxy
models -- the shape of the circular-speed curve, the values of the Oort
constants, etc -- relates almost exclusively to the radial force within the
plane. Such information is in principle incapable of determining how much of
the Galaxy's mass lies near the plane, which is clearly of prime importance
astrophysically.

This deficit of kinematic information has been papered over in two ways. The
first is a one-dimensional analysis of the vertical structure of the disk
along the lines pioneered by Oort (1932). Such analyses ultimately come up
against the problem that the vertical and horizontal motions of stars do not
decouple to the necessary degree, so that a one-dimensional analysis cannot
precisely determine the vertical distribution of matter -- see, e.g.,
Kuijken \& Gilmore (1991).

The second way in which the deficit of kinematic constraints on the Galaxy's
vertical structure has been papered over is to use star counts to constrain
the flattening of the components into which the overall luminosity density
is decomposed, and to assume that each component $i$ is characterized by a
constant mass-to-light ratio $\Upsilon_i$. This procedure is
methodologically questionable since it appears to presume that the
phenomenon of ``dark matter'' implies the existence of a completely dark
component comprised of exotic particles, whereas it may merely reflect the
variation from point to point of each $\Upsilon_i$. 

This is the first of a series of papers in which we plan to overcome these
difficulties by treating the orbits of stars in the meridional plane with
sufficient sophistication. Our approach, which has been described elsewhere
(Binney 1994, Dehnen \& Binney 1996), is iterative: we choose a potential,
determine a range of orbits in this potential, populate these with stars of
various spectral types and then compare the resulting predictions with the
available surveys. The potential is modified in the light of this
comparison. Thus our first step is to choose potentials (and thus,
implicitly, mass models) which are compatible with all the standard
kinematic and photometric constraints. This task of updating the Caldwell \&
Ostriker and Rohlfs \& Kreitschmann models is even now by no means trivial, as 
is made apparent by recent divergent results (Gates, Gyuk \& Turner, 1995; 
Cowsik, Ratnam \& Bhattacharjee, 1996; Evans 1996). Therefore we believe it 
will be useful to present in this paper our initial mass models and the consider
actions upon which they are based. Computer programs for evaluating the density 
and potential of the models are available upon request to the authors. Although
there is now abundant evidence that the inner Galaxy is significantly 
non-axisymmetric, our model conforms to the traditional axisymmetric pattern 
because the success of axisymmetric models in accounting for observations in 
the 21-cm line of hydrogen at longitudes $l\ga 30\deg$ suggests that orbits 
in the Galactic potential that carry stars to radii $r\ga 5\kpc$ can be 
accurately modelled by orbits in an axisymmetric potential.

We do not distinguish between the visible halo, of which RR-Lyrae
stars and metal-poor globular clusters are classical tracers, and the putative
dark halo: since we do not understand why the mass-to-light ratio rises with
galactocentric radius $r$, we are at liberty to assume that the Galaxy
possesses a single, massive halo that simply becomes more luminous with
decreasing $r$.

\section{FUNCTIONAL FORMS}	\label{sec-massmodel}
Our mass model contains three principal components: the disk, the bulge and
the halo. 

\subsection{The Disk} \label{sec-massmodel-disk}
Our disk is  made up of three components, namely the ISM, and the thin
and thick stellar disks. The density of each sub-disk is given by
 \beq \label{rho-disk}
	\rho_\d(R,z) = \frac{\Sigma_\d}{2 z_\d}
	\exp\left(-\frac{R_m}{R}-\frac{R}{R_\d}-\frac{|z|}{z_\d}\right),
\eeq With $R_m=0$, equation (\ref{rho-disk}) describes a standard double
exponential disk with scale-length $R_\d$, scale-height $z_\d$ and central
surface-density $\Sigma_d$. Since there appears to be very little
interstellar gas between the molecular ring at $R=4$-$5\kpc$ and the nuclear
disk at $\la200\parc$ (Dame \etal\ 1987), there should be a depression in
the central surface-density of the ISM. The parameter $R_m$ in equation
(\ref{rho-disk}) allows for such a central depression. We set $R_m=0$ for
the stellar disks and adopt $R_m=4\kpc$ for the ISM.  The total mass of a
disk with density (\ref{rho-disk}) is
 \beq \label{mass-disk}
	M_\d = 4 \upi\, \Sigma_\d\, R_m\, R_\d\, K_2(2\sqrt{R_m / R_\d}),
\eeq
 where $K_2$ is a modified Bessel function. For $R_m=0$ this gives 
$M_\d=2\upi\Sigma_\d\,R_\d^2$. 

Table \ref{tab-disk} gives our adopted values of the scale heights of the
three sub-disks as well as the fraction of the whole disk's surface density
at $R_0$ which is contributed by each sub-disk. As can also be seen from this
table, we fix the ratios between the sub-disk's scale-lengths. The value of
$R_{\d,\*}$, the scale length of the stellar disk, and the mass of the whole
disk are obtained from least-squared fits to the observational constraints to
be discussed below.

\begin{table}
	\caption[]{Fixed parameters of the disk components}
	\label{tab-disk}
\begin{tabular}{lp{2cm}rrr}
Component & Contribution
	    to $\Sigma(R_0)$& $R_\d/R_{\d,*}$ & $R_m$   & $z_\d$  \\ \hline
ISM	  & 0.25   	    & 2		      & $4\kpc$ &\hfil   40$\parc$\\
thin disk & 0.70   	    & 1		      & 0       &\hfil  180$\parc$\\
thick disk& 0.05   	    & 1		      & 0       &\hfil 1000$\parc$\\
\hline
\end{tabular}
\end{table}

\subsection{Bulge and Halo} \label{sec-massmodel-bulge+halo}
The bulge and halo are each described by the spheroidal density distribution
 \beq\label{rho-spheroid}
	\rho_\rmn{s} = \rho_0 \left(\frac{m}{r_0}\right)^{-\gamma}
			\left(1+\frac{m}{r_0}\right)^{\gamma-\beta}
			\rmn{e}^{-m^2/r_t^2},
\eeq
 where
\beq\label{ell-radius}
	m\equiv (R^2 + q^{-2}z^2)^{1/2}.
\eeq
 Thus the density of bulge and halo is proportional to $r^{-\gamma}$ for
$r\ll r_0$, proportional to $r^{-\beta}$ for $r_0\ll r\ll r_c$, and is
softly truncated at $r=r_t$. Infrared photometry obtained by the {\it 
COBE/DIRBE\/} satellite and analyzed by Spergel, Malhotra \& Blitz (1997)
yields values for four of the five bulge parameters: we adopt $\beta_\rmn{b}=
\gamma_\rmn{b}=1.8$, $q_\rmn{b}=0.6$, $r_{0,\rmn{b}}=1\kpc$, and 
$r_{t,\rm b}=1.9\kpc$. The density normalization $\rho_{0,\rm b}$, which is
not determined by the {\it COBE/DIRBE\/} data, is obtained from our
least-squared fits.

 The axis ratio of the halo is not significantly constrained by the 
observations discussed below, and we arbritraily set it to $q_{\rm h}=0.8$
(our standard value). However, we will also consider models with $q_{\rm h}=0.3$
motivated by (i) possible evidence that the halos of external galaxies might be
that flat \cite{oll96,sac94}, and (ii) Sciama's (1990) decaying neutrino model
for dark matter that requires a high local neutrino density
($\sim0.04\Msun\parc^{-2}$) only achievable with a flat neutrino halo.

 The remaining five halo parameters, $\beta_\rmn{h}$, $\gamma_\rmn{h}$,
$\rho_{0,\rm h}$, $r_{0,\rm h}$, and $r_{t,\rm h}$, are determined by the
least-squares fitting procedure described below subject to the restrictions
$-2\le \gamma_\rmn{h}\le\gamma_\rmn{b}$ and $\beta_\rmn{h}\ge 1$,
which limit the sharpness of any inner and outer edges of the halo,
respectively.

\subsection{Gravitational potential and forces}
	\label{sec-massmodel-potential}
The total gravitational potential $\Phi$ of a model must satisfy Poisson's
equation 
\beq\label{Poisson}
	\frac{\bmath{\nabla}^2\Phi}{4\upi\, G} =
	\sum_{j=1}^2 \rho_{\rmn{s},j} +
	\sum_{i=1}^3 \rho_{\rmn{d},i}.
\eeq
 A standard way to solve (\ref{Poisson}) involves expanding $\rho$ 
in spherical harmonics (c.f.\ BT section 2.8). Unfortunately, the
expansion of a thin disk  converges very slowly, so this straightforward
approach does not yield a fast and accurate solution of Poisson's equation.

 Fortunately, Kuijken \& Dubinski (1994) have described a modified multipole
technique that works well when the density is sum over components that are 
separable in cylindrical coordinates, that is, are of the form
\beq\label{rho-disk-factor}
	\rho_{\rmn{d},i} = f_i(R) \, h_i(z),
\eeq
 where $1=\int_{-\infty}^{\infty} h_i\, \D z$ so that $f(R)$ is the
radial surface-density profile.  Let $H_i(z)$ be such that
 \beq\label{defsH}
\begin{array}{r@{\quad}rl}
 \hbox{(i)}	   & H_i^{\prime\prime}(z) & \equiv h_i(z)		\\[1ex]
 \hbox{and\, (ii)} & H_i(0) 		   & =      H_i^\prime(0) = 0,
\end{array}
\eeq
 where the prime denotes a derivative as usual. Then we write
 \beq\label{pot-ansatz}
	\Phi(R,z) = \Phi_{ME}(R,z) + 4\upi\, G \sum_i f_i(r) H_i(z),
\eeq
 where the argument of $f_i$ is now the spherical radius $r$ rather than the
cylindrical radius $R$ and $\Phi_{ME}$ is a function to be determined. At
$z=0$ the second term on the right hand side of (\ref{pot-ansatz}) and its
first derivatives vanish, so both the potential and the forces in the plane
are determined by $\Phi_{ME}$ alone.  Inserting (\ref{rho-disk-factor}) and
(\ref{pot-ansatz}) into (\ref{Poisson}) we obtain for $\Phi_{ME}$
 \beq\label{KDeq}
\begin{array}{rl}
	  \dsy{\frac{\bmath{\nabla}^2\Phi_{ME}}{4\upi\, G} = } &
	  \dsy{\sum_j \rho_{\rmn{s},j} +
	      \sum_i \bigg\{ \big[f_i(R)-f_i(r)\big]h_i(z)} \\
	& \dsy{-f_i^{\prime\prime}(r)H_i(z)-\frac{2}{r}f_i^\prime(r)
		               \big[H_i(z)+zH_i^\prime(z)\big]\bigg\}}.
\end{array}
\eeq
 This equation takes the form of Poisson's equation for $\Phi_{ME}$ with a
mass-density given by the complex expression on its right-hand side. At
$z=0$ we have $R=r$, so with (\ref{defsH}) this expression simplifies to
$\sum_j\rho_{\rmn{s},j}$. That is, $\Phi_{ME}$ is generated by a mass 
distribution that is not strongly confined to the plane, and can be
economically evaluated by expanding both sides of equation (\ref{KDeq}) in
spherical harmonics.

\begin{table}
	\caption[]{Pairs of functions $h(z)$ and $H(z)$ defined in section
		   \ref{sec-massmodel-potential}.} \label{tab-h-pair}
\begin{tabular}{l@{\hspace{1cm}}l}
	$h(z)\equiv \D^2H/\D z^2 $ \hfill	&
	$H(z) $ \hfill				\\ \hline
	$\delta(z)$ \hfill			&
	$\frac{|z|}{2}$	 \hfill		\\[2ex]
	$\frac{1}{2z_\d}\,\exp(-\frac{|z|}{z_\d})$ \hfill	&
	$\frac{z_\d}{2}\left[\exp(-\frac{|z|}{z_\d})-1+\frac{|z|}{z_\d}\right]$
						\hfill	\\[2ex]
	$\frac{1}{4z_\d}\sech^2\frac{z}{2z_\d} $	\hfill		&
	$z_\d\,\ln\cosh\frac{z}{2z_\d} $	\hfill		\\ \hline
\end{tabular}
\end{table}

Once $\Phi_{ME}$ has been found, it and its first derivatives are stored on
a grid in $\ln r$ and $|z/r|$. A two-dimensional fifth-order spline is used
to interpolate on this grid: at each grid point this spline yields the
stored values of the potential and its derivatives, and the forces have
everywhere continuous first and second derivatives. In particular, the
interpolated forces agree with the derivatives of the interpolated
potential, as is necessary if energy is to accurately conserved along
numerically integrated orbits. Furthermore, the evaluation of potential and
forces is quick once the spline coefficients have been computed.

We plan to make available after publication of this paper a {\sf C++} source
code for the evaluation of the potential of any superposition of spheroids
(eq.\ \ref{rho-spheroid}) and exponential disks with vertical profiles as
in Table \ref{tab-h-pair} -- send e-mail to w.dehnen\symbol{64}physics.ox.ac.uk.
(Public-domain {\sf C} compilers are available that allow {\sf C++} code to be
linked to otherwise pure {\sf C} programs.)

\section{THE OBSERVATIONAL CONSTRAINTS} \label{sec-constraints}
Three groups of observational data  constrain the values of the free
parameters in the model introduced in section \ref{sec-massmodel}: (i) tangent
velocities at $R<R_0$; (ii) rotation velocities at $R>R_0$; (iii) other
data, such as the values of Oort's constants and the local surface density.
We discuss each group of constraints separately.

\subsection{Terminal velocities for the inner Galaxy}
	\label{sec-constraint-inner}
For an axisymmetric Galaxy with circularly rotating ISM, the peak velocity
along a given line-of-sight at $b=0$ and $l$ in either the first or fourth
quadrant originates from the radius $R=R_0\sin l$. Relative to the LSR this
`terminal velocity' is related to the circular speed $\vc$ by
 \beq \label{inner-vtangent}
	v_\rmn{term} = \vc(R_0\,\sin l) - \vc(R_0)\,\sin l.
\eeq
 In reality, non-circular motions of the ISM induced, for instance, by
spiral arms, lead to deviations from this ideal relation. However, outside
the region $|l|\la20\deg$ that is dominated by the bar, these deviations
are expected to be ignorable for our purposes.

Numerous surveys of the ISM have been undertaken. In this study we restrict
ourselves to three surveys in \HI\ \cite{ww73,bl84,kea86} and one in CO
\cite{ksw85}. Malhotra (1994,1995) has modelled these raw data in detail and
kindly provided us her values for the terminal velocities in electronically
readable form. We restrict the data to $|\sin l|\ge 0.3$ to avoid distortions
by the central bar.

\subsection{The rotation curve of the outer Galaxy}
	\label{sec-constraint-outer}
For an axisymmetric galaxy, the radial velocity relative to the LSR,
$v_\rmn{lsr}$, of a circularly orbiting object at galactic coordinates $(l,b)$
and galactocentric radius $R$ is related to the circular speed by
\beq \label{outer-vlsr}
	W(R) \equiv \frac{v_\rmn{lsr}}{\sin l\,\cos b}
	     =      \frac{R_0}{R}\vc(R) - \vc(R_0).
\eeq 
 As is well known, for $R>R_0$ one cannot infer $R$ for an object at given
$l$ without a knowledge of the distance $d$ to the object. If $d$ is known,
then $R$ follows from
 \beq \label{outer-radius}
	R = \left( d^2\cos^2\!b + R_0^2 - 2 R_0 d\, \cos b\,\cos l\right)^{1/2}.
\eeq

 Several studies are available that contain measured values of $W$\/ and $d$
for objects that ought to be on nearly circular orbits. Here we use the data
of two recent studies. Brand \& Blitz (1993) list \HII\ regions/reflection
nebulae that have (spectro-) photometric distances and associated molecular
clouds with measured radial velocities. Pont~et~al.\ (1997) give radial
velocities and photometry for classical cepheids in the outer disk. For
these objects we have
transformed $v_r$ to $v_\rmn{lsr}$ and evaluated the distances using the
period-luminosity relation derived from Hipparcos paralaxes by Feast and
co-workers \cite{FC97,FW97ab}, $M_V=-2.81 \log P - 1.43$, in conjunction with
a Galactic period-colour relation from Laney \& Stobie (1994):
$(B-V)_0 = 0.416 \log P + 0.314$. 

We have rejected objects for which either $155^\circ\le l\le 205^\circ$, or
$d<1\kpc$, or $W>0$, because for these objects $v_\rmn{lsr}$ is very likely
dominated by non-circular motions. Furthermore, we have not used data points
at $R < R_0$ where the terminal velocities provide a much better constraint.
93 of the 205 objects in Brand \& Blitz (1993) and 26 of the 48 Cepheids
survived this cull.

We have not employed a number of data sets that are similar in scope to
those of Brand \& Blitz and Feast \etal\ because these sets are either rather
restricted in their radial coverage, or have problematic distances. For
example, the distances to carbon stars are seriously affected by both
Malmquist bias and interstellar extinction (Schechter, private
communication).

A technique for measuring $W(R)$ from 21-cm emission without independent
distance information has been proposed by Merrifield (1992). This involves a
determination of the extent in $b$ of the emission observed at given $W$: in
an axisymmetric galaxy with circularly orbiting \HI, all emission at given
$W$ will originate in a galactocentric ring. If this ring has a constant
vertical extension, it creates a distinct pattern in the $(l,b)$ plane. From
the \HI\ surveys of Weaver \& Williams (1974) and Kerr \etal\ (1986)
Merrifield estimated relative galactocentric distances $R/R_0$ by fitting
this characteristic pattern to the emission from each bin in $W$. It is not
immediately apparent how accurate Merrifield's radii are since both random
motions in the plane and systematic variations in the thickness of the \HI\
layer around circles will contribute errors.

\ifREFEREE \relax
\else
\begin{figure}
        \epsfxsize=21pc \epsfbox[18 170 600 700]{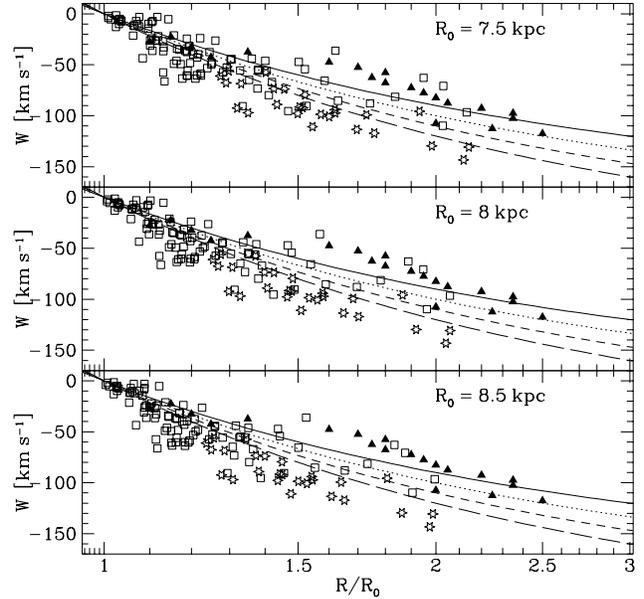}
 \caption[]{Data on the outer rotation curve: $W$ vs.\ $R/R_0$ (errorbars are
	    omitted for clarity). Squares, stars, and triangles represent 
 	    \HII\ regions, Cepheids, and \HI\ measurements, respectively.
 	    The lines refer to flat rotation curves with $v_c(R_0)=180\kms$
	    (solid), 200\kms\ (dotted), 220\kms\ (short dashed), and 240\kms\
	    (long dashed), respectively. Each panel corresponds to a different
	    assumed value of $R_0$; increasing $R_0$ shifts the squares and the 
	    stars to the left.}
 \label{fig-W-data}
\end{figure}
\fi

It is worth comparing the three data sets before trying to fit them to a
mass model. Fig.~\ref{fig-W-data} shows the data in $W$ vs.\ $R/R_0$ for
$R_0=7.5\kpc$, 8\kpc, and 8.5\kpc. Also shown is $W(R)$ predicted by flat
rotation curves for $v_c(R_0)=180\kms$ (uppermost line), 200\kms, 220\kms,
and 240\kms\ (lowest line). Several points can be drawn immediately from
this Figure.  (i) The data of Merrifield (using \HI) at $R>1.3R_0$ are
incompatible with the Cepheid data \cite{Pontetal97}. (ii) The \HII\ data at
$R<1.5R_0$ and the Cepheid data imply a falling rotation curve, unless
$v_c(R_0)\ga240\kms$ and/or $R_0\la7.5\kpc$. (iii) Merrifield's measurements
require a rising rotation curve, unless $v_c(R_0)\la 180\kms$.

Clearly, it
makes no sense trying to fit inconsistent data; we must decide which
data to trust. Cepheid distances have been extensively studied, whereas
Merrifield's novel method rests on several ad hoc assumptions. Furthermore,
his values for $W$ are suspect because they are non-monotonic at $R>2R_0$. (Note
that they do not represent individual objects as do the other data points.)
Therefore, we have decided to discard Merrifield's points altogether.

\subsection{Other constraints}
	\label{sec-constraint-local+global}
 By dividing equation (\ref{inner-vtangent}) by $R_0\sin l$ and equation
(\ref{outer-vlsr}) by $R_0$, one sees that studies of the ISM measure
$\Omega\equiv\vc(R)/R$ at various positions in the Milky Way {\it relative\/}
to its local value. To fix the {\it absolute\/} values of $\vc$, additional
information is essential.

\subsubsection{Oort's constants}
  Oort's constants $A$ and $B$ are defined by
\beq \label{oortsAB-axi}
\begin{array}{lll}
	A & =   & \frac{1}{2}\dsy\left({\vc\over R}-{\p\vc \over \p R}\right)
	\\[0.5ex]
	B & = - & \frac{1}{2}\dsy\left({\vc\over R}+{\p\vc \over \p R}\right).
\end{array}
\eeq
 They can be derived from the kinematics of nearby stars. Kerr \& Lynden-Bell 
(1986) reviewed the published measurements and concluded that
			$A  = 14.4\pm1.2$,
			$B  =-12.0\pm2.8$, and
			$A-B= 26.4\pm1.9$,
all in units of $\kms\kpc^{-1}$. Hanson (1987) found from an analysis of
the NPM catalog 	$A  = 11.3\pm1.1$,
			$B  =-13.9\pm0.9$, and
			$A-B= 25.2\pm1.6$, while a more recent
study of Hipparcos proper motions for Cepheids by Feast \& Whitelock (1997b)
yields			$A   =14.8\pm0.8 $, and
			$A-B =27.2\pm0.9 $ in the same units.
Since the Hipparcos values are likely to be significantly more reliable than
the earlier studies, we give them the highest weight. 

A potentially relevant measurement is of the proper motion of \SgrA:
$\mu_l=(-6.55\pm0.17)\masyr=(-31.05\pm0.81)\kms\kpc^{-1}$ \cite{bac96}. To
this the peculiar motion of the Sun contributes
$V_\odot/R_0\simeq1\kms\kpc^{-1}$. Hence, if we believe that \SgrA, which is
certainly massive \cite{Genzeletal}, is stationary at the centre of the
Galaxy then we have $A-B=30\pm1\kms\kpc^{-1}$, which is significantly
inconsistent with the result of Feast \& Whitelock. In view of this conflict
and the possibility that the Galactic centre is oscillating, perhaps in an
$m=1$ mode, (Gould \& Ram\'{\i}rez 1997) we have not used Backer's result
but imposed the constraints\footnote{The product $AR_0$ is not an
independent observable, as it is usually measured from the ISM's terminal
velocities, which have already been employed as constraints.}
\beq 
 \label{local-oort}
 \begin{array}{rcclclrl}

 A   & = &   & 14.5 & \pm & 1.5 & \kms\kpc^{-1} & \\
 B   & = & - & 12.5 & \pm & 2   & \kms\kpc^{-1} & \\
 A-B & = &   & 27   & \pm & 1.5 & \kms\kpc^{-1} & .
 \end{array}
\eeq

\subsubsection{The mass at large radii}
While observations for $A$, $B$, and $v_c(R_0)$ restrict the circular speed
locally, and hence the mass inside $R_0$, there are some important 
constraints at much larger radii. The total mass within a sphere of radius
$r\gg R_0$ can be estimated
(i)   from the velocity distribution of the Milky Way's satellites,
(ii)  from the maximal locally observed stellar velocity (the `escape velocity'
      argument),
(iii) the timing of the local group, and
(iv)  by modelling the dynamics of the Magellanic Clouds and Stream.
 All these estimates rely on certain assumptions and are model dependent.
However, with reasonable assumptions Kochanek (1996) found a simple model
that satisfies the first three of these constraints and yields an acceptable
value of $v_c(R_0)$. From his Figure 7 we extracted for the mass inside
100\kpc\ $M_{R<100\rmn{kpc}}=7.5\pm2\times 10^{11}\Msun$. For comparison, by
modelling the dynamics of the Magellanic Clouds and Stream Lin, Jones \&
Klemola (1995) found that $M_{R<100\rmn{kpc}}=5.5\pm1\times 10^{11}\Msun$.
Clearly, the uncertainties here are dominated by systematic errors, so we
allow for a generous error bound on our adopted constraint, which is
 \beq \label{global-M}
	M_{R<100\rmn{kpc}}=(7\pm2.5)\times 10^{11}\Msun.
\eeq

\subsubsection{The local vertical force}
The vertical force $K_z$ at some height above the plane places a condition
on the local mass distribution, and certainly is an important observable our
model must agree with. Using K stars as a tracer population, Kuijken \&
Gilmore (1989,1991) have deduced
 \beq \label{local-Kz}
	K_{z,{}_{1.1}} \equiv
	|K_z (R_0,1.1\kpc)| = 2\upi\, G \times (71\pm6) \Msun\parc^{-2}.
\eeq
 We have adopted this as a constraint for our models.

\subsubsection{The disk's local surface density}
 Unfortunately, the local disk surface
density $\Sigma_0$  is not as well determined as the closely related quantity
$K_{z,{}_{1.1}}$ \cite{kg91}. However, by counting identified matter Kuijken
\& Gilmore (1989) concluded that $\Sigma_\rmn{stars+gas}(R_0)=48\pm8\,\Msun
\parc^{-2}$. We adopt the constraint
\beq \label{local-sigma-nod}
	\Sigma_0 \equiv \Sigma_\rmn{disk}(R_0) \ge 40 \, \Msun \parc^{-2}.
\eeq

\subsubsection{The dispersion velocity in Baade's window}
Finally, an important constraint on the bulge is provided by the observed
velocity dispersion of the bulge in Baade's window of $117\pm5\kms$
\cite{rich88,tsr95}. The simplest way to estimate the velocity dispersion of
our models is to solve the Jeans equations assuming isotropy in the velocities,
which yields
 \beq\label{Jeqest}
	\sigma^2_{\rm b} = \frac{1}{\rho_{\rm b}}
		     \int_z^\infty \rho_{\rm b}\, \frac{\p\Phi}{\p z}\, \D z,
\eeq
 where the subscript b stands for bulge. The bulge is now known to be
significantly elongated towards us (e.g., Binney, Gerhard \& Spergel, 1997)
and this elongation is probably reflected in the line-of-sight dispersion
along the Galaxy's minor axis being larger than equation (\ref{Jeqest})
allows. On the other hand, the velocity dispersion probably falls as one
moves away from the minor axis along the line of sight, and this effect will
tend to cause equation (\ref{Jeqest}) to overestimate the measured
dispersion within Baade's window.  In view of these oppositely directed
factors, we adopt (\ref{Jeqest}) as the central value of our constraint on
the dispersion within Baade's window and allow a wide range around this
value:
 \beq \label{global-Baade}
	\sigma_{{}_{BW}}\equiv
	\sigma_{\rm b}(0.0175R_0,-0.068R_0)
	=117\pm15\kms.
\eeq

\section{FITTING THE MASS MODEL} \label{sec-fitting}
The free parameters of the mass model described in section \ref{sec-massmodel} 
are determined by minimizing the quantity
\beq \label{chisq-tot}
\chi^2_{{}_\rmn{tot}} = 
	  \frac{W_{{}_\rmn{in}}}{N_{{}_\rmn{in}}}\chi^2_{{}_\rmn{in}}
	+ \frac{W_{{}_\rmn{out}}}{N_{{}_\rmn{out}}}\chi^2_{{}_\rmn{out}} 
	+ \frac{W_{{}_\rmn{other}}}{N_{{}_\rmn{other}}}\chi^2_{{}_\rmn{other}} 
\eeq
that is the sum of pseudo-chi-squared contributions from our three classes of 
constraint. Here the $N_i$ are the numbers of data points actually used, while
the $W_i$ are weights, which may be interpreted as the number of really
independent constraints (for instance, Oort's $A$ has been obtained from much
more data than we use for $v_\rmn{term}$). Clearly, the $W_i$ are subject to 
ones prejudices, we took
$W_{{}_\rmn{in}}=W_{{}_\rmn{out}}=W_{{}_\rmn{other}}=N_{{}_\rmn{other}}=6$.

There are contributions to $\chi^2_{{}_\rmn{in}}$ from 53 data points at
$l<0$ and 77 at $l>0$. In order to minimize the influence of
systematic deviations from circular motion, which differ on the two sides of
the Galaxy, the data for positive and negative longitude are weighted by
0.844 and 1.23, respectively.  This gives an effective number of 65 data
points on either side, while leaving the effective total number of data points
unchanged. In order to allow for non-circular motions both random and
systematic, we adopt a constant uncertainty of 7\kms\ for $v_\rmn{term}$.
Hence each data point adds to  $\chi^2_{{}_\rmn{in}}$ an amount
 \beq
w\left({v_\rmn{term,\,model}(l)-v_\rmn{term,\,data}(l)\over7\kms}\right)^2
\eeq
where $w=0.844$ or $w=1.23$ depending on whether $l$ is greater than or
less than 0.

The rotation-curve data for $R>R_0$ cannot be treated in an exactly
analogous way because now two numbers contain significant uncertainties: $d$
and $W$.  Following Fich, Blitz \& Stark (1989) we take the
contribution to $\chi^2_{{}_\rmn{out}}$ from the $i$th data point to be
 \beq \label{chisq-generalized}
	w\min_{R>R_0} \left(
	\left[\frac{\ln d_i - \ln d_\rmn{model}(R)}{\Delta \ln d_i}\right]^2 +
	\left[\frac{W_i - W_\rmn{model}(R)}{\Delta W_i}\right]^2   \right),
\eeq
 where $\Delta\ln d$ and $\Delta W$ are the uncertainties in the
observables. In order to give each catalog the same weight, the data taken
from Brand \& Blitz (1993) and Pont \etal\ (1997) are weighted by $w=0.665$
and $2.016$, respectively, leaving the effective total number of data points
unchanged at 119. To account for non-circular motions, a dispersion velocity
of 7\kms\ is quadratically added to the measurement errors of $v_\rmn{lsr}$.

\begin{table*}
\begin{minipage}{175mm}
\tabcolsep1mm
\caption[]{Fit results}\label{tab-fit}
 \def\telf{$\times10^{11}$}
 \begin{tabular}{cclccccccccrrrr}
                %123456789 1234567
 Model   & $\frac{R_{\rmn{d},*}}{R_0}$ & Deviations from standard &
         $\Sigma_0$ &$A$ &$B$ &$v_c(R_0)$ & $\frac{K_{z,1.1}}{2\upi G}$
         &$M_{R<100\rmn{kpc}}$ &$\sigma_{{}_{BW}}$ &
         $\chi^2_{{}_\rmn{in}}$  &$\chi^2_{{}_\rmn{out}}$ &
         $\chi^2_{{}_\rmn{other}}$ &$\chi^2_{{}_\rmn{tot}}$      \\ \hline
1 &0.25&--&43.3&14.4&--13.3&222&68.0&6.56\telf&100 & 91.2&174&1.92 & 13.9 \\%
2 &0.30&--&52.1&14.3&--12.9&217&72.2&6.52\telf&108 & 82.3&171&0.53 & 12.2 \\%
3 &0.35&--&52.7&14.1&--13.1&217&72.5&6.51\telf&111 & 86.7&175&0.46 & 12.5 \\%
4 &0.40&--&50.7&13.8&--13.6&220&72.1&6.04\telf&111 & 93.7&180&0.94 & 13.5 \\%
[1ex]%
2a&0.30& $R_0=7.5\kpc$
        &52.5&15.1&--12.2&204&71.3&6.95\telf&106 & 83.5&164&0.70 & 12.1 \\%
2b&0.30& $R_0=8.5\kpc$
        &50.3&13.8&--13.3&231&71.9&6.33\telf&108 & 81.0&176&0.87 & 12.7 \\%
2c&0.30& $\gamma_\rmn{h}=1$
        &52.3&14.3&--12.7&216&72.0&6.77\telf&105 & 83.9&171&0.69 & 12.4 \\%
2d&0.30& $\gamma_\rmn{h}=1$ and $\beta_\rmn{h}=3$
        &53.3&14.2&--12.9&217&72.9&6.33\telf&108 & 84.3&176&0.66 & 12.7 \\%
2e&0.30& $0.1 \cos R/R_\rmn{d}$ added to exponent in (1)
        &49.6&14.3&--13.0&218&71.3&6.49\telf&105 & 85.5&176&0.78 & 12.8 \\%
2f&0.30& $-0.1 \cos R/R_\rmn{d}$ added to exponent in (1)
        &52.8&14.4&--12.6&217&72.0&6.77\telf&108 & 80.6&166&0.42 & 11.8 \\%
2g&0.30& constraint $\sigma_{BW}=140\pm15\kms$
        &51.4&14.1&--13.2&218&72.0&6.39\telf&115 & 95.1&176&2.99 & 15.1 \\%
2h&0.30& $M_\rmn{b}\ge1.5\times10^{10}\Msun$
        &44.1&13.9&--14.0&223&68.5&6.01\telf&124 & 131 &181&1.71 & 15.9 \\%
2i&0.30& halo axis ratio $q_\rmn{h}=0.3$
        &40.0&14.0&--11.9&207&77.9&5.59\telf&109 & 88.8&187&2.68 & 15.1 \\%
[1ex]%
4a&0.40& $R_0=7.5\kpc$
        &49.6&14.3&--13.6&210&71.5&5.43\telf&116 & 103 &175&1.16 & 13.9 \\%
4b&0.40& $R_0=8.5\kpc$
        &50.8&13.4&--13.7&231&72.4&6.29\telf&107 & 89.1&186&1.50 & 14.1 \\%
4c&0.40& $\gamma_\rmn{h}=1$
        &49.8&13.9&--13.7&221&71.9&5.75\telf&106 & 92.1&178&1.48 & 13.8 \\%
4d&0.40& $\gamma_\rmn{h}=1$ and $\beta_\rmn{h}=3$
        &48.3&14.0&--13.8&222&71.4&3.73\telf&107 & 89.6&173&2.99 & 14.7 \\%
4e&0.40& $0.1 \cos R/R_\rmn{d}$ added to exponent in (1)
        &51.7&14.0&--13.3&218&72.5&6.36\telf&111 & 88.6&179&0.57 & 12.9 \\%
4d&0.40& $-0.1 \cos R/R_\rmn{d}$ added to exponent in (1)
        &49.5&13.7&--13.9&221&71.7&5.76\telf&110 & 99.1&182&1.41 & 14.2 \\%
4g&0.40& constraint $\sigma_{BW}=140\pm15\kms$
        &52.0&13.8&--13.4&218&72.4&6.27\telf&117 & 106 &182&2.87 & 15.8 \\%
4h&0.40& $M_\rmn{b}\ge1.5\times10^{10}\Msun$
        &48.3&13.7&--14.2&223&71.5&4.89\telf&120 & 114 &181&2.06 & 15.5 \\%
4i&0.40& halo axis ratio $q_\rmn{h}=0.3$
        &40.0&13.6&--12.3&207&80.2&3.41\telf&112 & 105 &197&5.57 & 18.7 \\%
 \hline
 \end{tabular}\par\medskip

The standard models 1 to 4 are determined by the choice of $R_{\d,*}/R_0$
(column 3), where $R_0=8\kpc$. The non-standard models 2a-2i and 4a-4i are
specified in column 4. Columns 5 to 11 give the best-fit values of
the observables discussed in Section \ref{sec-constraint-local+global}. The
values obtained for the $\chi^2$s defined in Section \ref{sec-fitting} are
given in the last four columns. The units are as usual:
    $R_0$ in kpc;
    $\Sigma_0$, $K_{z,1.1}/(2\upi G)$ in $\rmn{M}_\odot\rmn{pc}^{-2}$;
    $A$, $B$ in $\rmn{km\,s}^{-1}\rmn{kpc}^{-1}$; $M_{R<100\rmn{kpc}}$ in
    $\Msun$; and $v_c(R_0)$ and $\sigma_{{}_{BW}}$ in $\rmn{km\,s}^{-1}$.
\end{minipage}
\end{table*}

\ifREFEREE \relax
\else
\begin{figure}
        \epsfxsize=21pc \epsfbox[18 150 460 700]{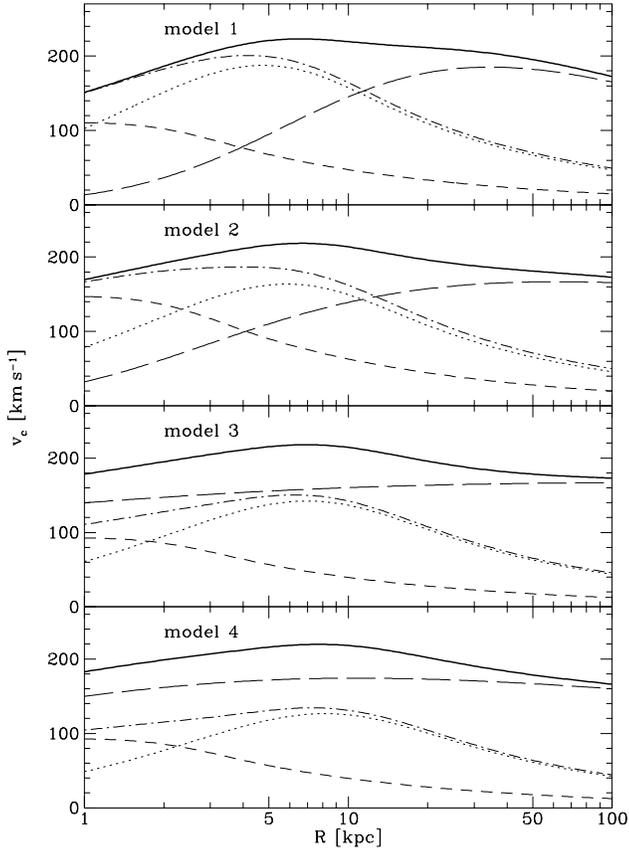}
 \caption[]{Rotation curves of models 1-4 (thick solid lines). The circular 
	velocities due the disk (dotted), bulge (short dashed), disk and bulge
	(dash dotted), and halo (long dashed) are also shown.}
 \label{fig-vc-standard}
\end{figure}

\begin{figure}
        \epsfxsize=21pc \epsfbox[18 250 620 700]{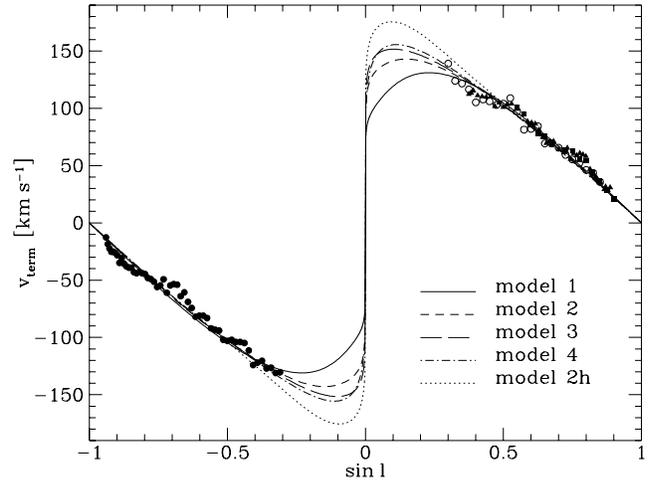}
 \caption[]{Terminal velocities at $R\le R_0$ for models 1-4 and 2h.
	The data are from Weaver \& Williams (1973, filled triangles), Bania \&
	Lockman (1984, filled squares), Knapp \etal (1985, open circles), and
	Kerr \etal (1986, filled circles).}
 \label{fig-vt-standard}
\end{figure}

\begin{figure}
        \epsfxsize=21pc \epsfbox[20 270 620 700]{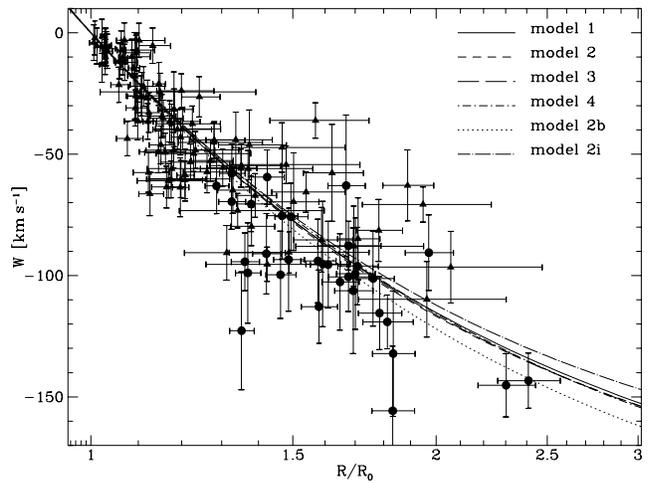}
 \caption[]{$W$ (equation \ref{outer-vlsr}) versus $R/R_0$ for models 1-4
	(hardly to distinguish) and the two most extreme models 2b,i. The data
	are from Brand \& Blitz (1993; triangles) and Pont et~al. (1997; dots).}
 \label{fig-W-standard}
\end{figure}
\fi

\section{RESULTS}\label{sec-results}
There are two aspects of any given model to consider: (i) how well does it
fit the observational constraints, and (ii) how is its mass distributed.
Since the observational constraints mostly relate to motions in the plane,
these two questions are in large degree independent of one another. Table
\ref{tab-fit}  lists the defining characteristics of each model and the
quality of the fit to the data that it furnishes. Table \ref{tab-best-fit}
lists the values of each model's parameters.

The most important parameter of the models proves to be $R_{\d,*}/R_0$.  The
first four rows of Tables \ref{tab-fit} and \ref{tab-best-fit} refer to our
`standard models', which all have $R_0=8\kpc$ but adopt four values of
$R_{\d,*}/R_0$: $0.25$, $0.3$, $0.35$ and $0.4$. Each model is determined by
specifying the value of $R_{\d,*}/R_0$ and then solving for the values of
the other model parameters which minimize $\chi^2_{{}_\rmn{tot}}$. These
values are given in Table
\ref{tab-best-fit}.

Fig.~\ref{fig-vc-standard} shows the circular-speed curves predicted by the
four standard models together with the contributions to
$\vc$ from each component. In all four models the circular speed declines
outside $R_0$, although in Model 4 the decline is extremely slow near $R_0$.

As $R_{\d,*}/R_0$ increases from $0.25$ to $0.4$, the peak in the disk's
contribution to $\vc$ moves outwards from $5\kpc$ and the amplitude of the
disk's contribution to $\vc$ declines markedly. This decline in the disk's
contribution to the inner circular-speed curve is compensated by an increase
in the halo's contribution. This increase is achieved by making the halo
more centrally concentrated, with the result that for $R_{\d,*}/R_0\ge0.35$
the halo does not differ greatly from a pure power-law component whose
conbtribution to $\vc$ is nearly independent of radius. For $R_{\d,*}/R_0=0.4$
this contribution dominates $\vc$ at {\em all\/} radii.

The parameter $\Sigma_\rmn{d,tot}$ is the sum of the values for the three
disks of the parameter $\Sigma_\rmn{d}$ that is defined by equation
(\ref{rho-disk}). Table \ref{tab-best-fit} shows that $\Sigma_\rmn{d,tot}$
decreases by a factor of almost 4 as $R_{\rmn{d},*}/R_0$ increases from
$0.25$ to $0.4$. Consequently, the disk's contribution to $v_c$ at small
radii decreases by a factor of 2, which is first compensated by a near doubling
in the bulge mass, and then by a dramatic increase in the central density of
the halo. Specifically, whereas for the smallest two values of $R_{\rmn{d},*}
/R_0$ the halo has a hole at its centre ($\gamma_\h=-2$), for all larger values
of $R_{\d,*}/R_0$ the halo density decreases outwards as $\sim r^{-1.7}$.

The amplitude of the bulge's contribution to $\vc$ near the centre is
largest for $R_{\d,*}/R_0=0.3$ -- smaller and larger values of this parameter
yield bulges that are less massive by a factor in excess of 40 per cent.
In all four models the velocity dispersion in Baade's window lies below the
target value. As we indicated above, this shortfall probably reflects the
fact that our models take no account of the elongation of the bulge along
the line of sight.

The model with the smallest value of $R_{\d,*}/R_0$ is nearly a maximum-disk
model in the sense that there is a wide range of radii $R\la10\kpc$ within
which nearly all of $\vc$ derives from the disk alone. In Model 2, which has
$R_{\d,*}/R_0=0.3$, the disk and bulge together dominate $\vc$ at
$R\la10\kpc$; the principal difference between Models 1 and 2 is the
dominance of the bulge at $R\la2\kpc$ discussed above. For $R_{\d,*}/R_0\ga
0.35$ the disk's constribution to $v_c$ is nowhere as important as the halo,
even though the disks of these models are only slightly less massive than those
of models with $R_{\d,*}/R_0\la0.3$.

All the standard models have masses $M_{R<100\rmn{kpc}}$ that lie below the 
target value. This is because the rotation-curve data for $R>R_0$ force $\vc$ to
decline just beyond the Sun. In extreme cases -- for example Model 3 and
Model 2a below -- the constraint on $M_{R<100\rmn{kpc}}$ obliges the halo to
make an outwards-increasing contribution to $\vc$ even at the largest radii,
with the result that $\vc$ is increasing at $R=100\kpc$.

All four standard models provide satisfactory fits to the constraints of
Section 3.3 -- see Table \ref{tab-fit}.  Fig.~\ref{fig-vt-standard} shows
that these models also provide excellent fits to the observed tangent
velocities at $R<R_0$: the deviations between observed and predicted points
are of the order of those expected to arise from spiral structure.
Fig.~\ref{fig-W-standard} shows that all four standard models provide 
essentially the same reasonable fit to the (widely scattered) rotation-curve
data for $R>R_0$.

\ifREFEREE \relax
\else
\begin{figure}
        \epsfxsize=21pc \epsfbox[18 150 460 700]{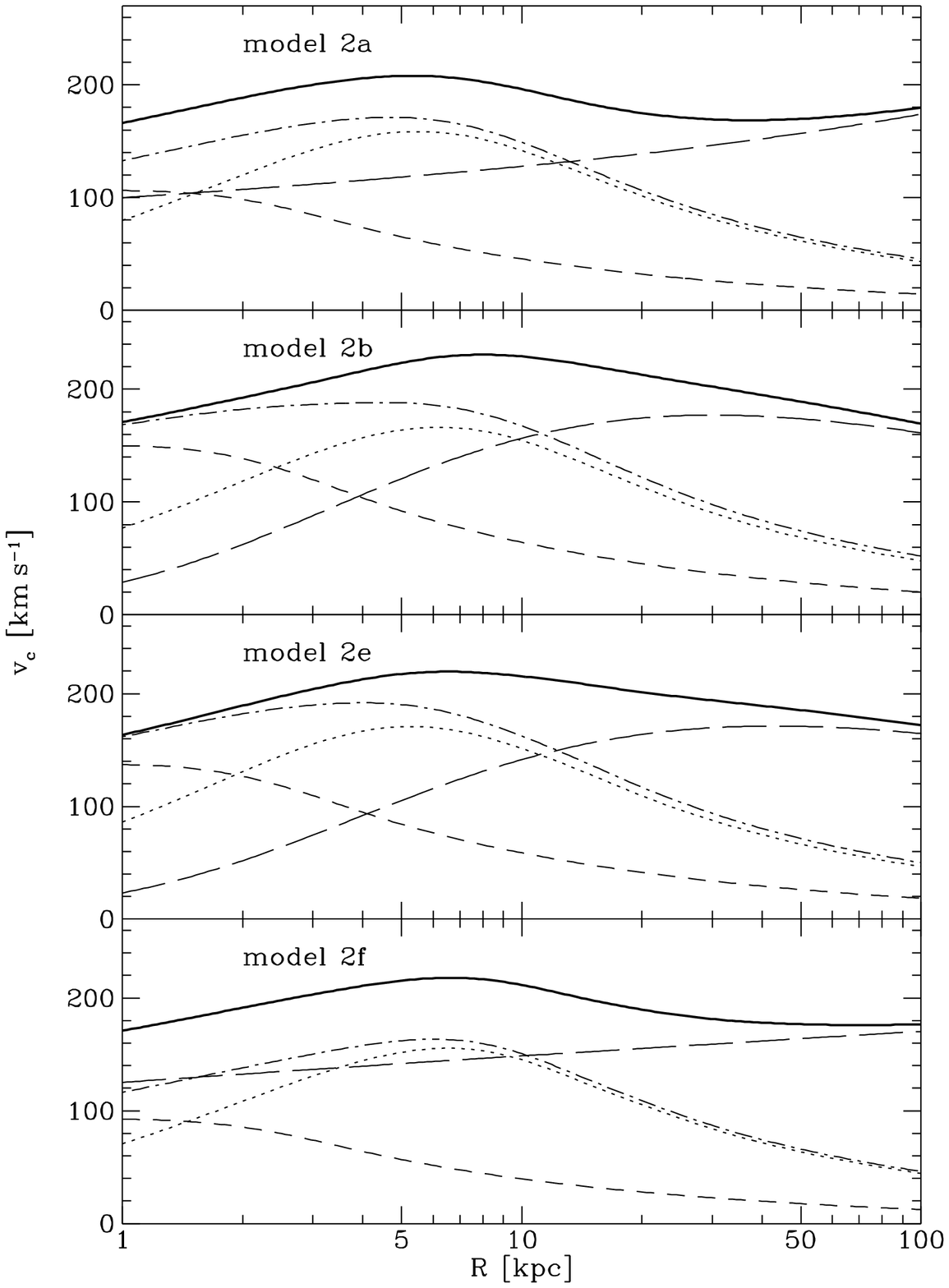}
 \caption[]{Rotation curves of models 2a,b,e,f (thick solid lines). The circular
	velocities due the disk (dotted), bulge (short dashed), disk and bulge
	(dash dotted), and halo (long dashed) are also shown.}
 \label{fig-vc-non-standard}
\end{figure}
\fi

\begin{table*}
\begin{minipage}{175mm}
\caption[]{Best-fit values of the model parameters}\label{tab-best-fit}
 \begin{tabular}{lccccccc @{\hspace{7mm}} % model + parameters
                 cccc}                    % masses
 Model   &$\Sigma_{\rmn{d,tot}}$ &$\rho_{0,b}$ &$\rho_{0,\rmn h}$ 
         &$\gamma_{\rmn h}$ & $\beta_{\rmn h}$ & $r_{0,\rmn h}$ 
         &$r_{\rmn t,h}$ & $M_d$ & $M_{\rm b}$ 
         &$M_{h,<10\rmn{kpc}}$ & $M_{h,<100\rmn{kpc}}$ \\ \hline
1 &1905&0.4271&0.7110&--2&2.959&3.83&$\infty$&5.13&0.518&2.81&60.0 \\%
2 &1208&0.7561&1.263&--2&2.207&1.09&$\infty$&4.88&0.917&2.89&59.4 \\%
3 &778.4&0.3&0.1179&1.8&2.002&2.29&$\infty$&4.46&0.364&4.36&60.3 \\%
4 &536&0.3&0.2659&1.629&2.167&1.899&$\infty$&4.16&0.364&5.23&55.9 \\%
[1ex]%
2a&1222&0.3958&0.0009166&1.8&1.634&22.79&$\infty$&4.32&0.48&2.46&64.7 \\%
2b&1165&0.7874&1.425&--2&2.553&1.733&$\infty$&5.32&0.955&4&57.1 \\%
2c&1213&0.5995&0.03055&1&2.209&6.185&$\infty$&4.9&0.727&2.99&62.1 \\%
2d&1238&0.6754&0.006159&1&3&21.8&$\infty$&5&0.819&2.83&57.5 \\%
2e&1260&0.6584&0.9084&--2&2.425&1.802&$\infty$&5.04&0.798&2.88&59 \\%
2f&1121&0.3&0.4179&1.8&1.888&1&$\infty$&4.6&0.364&3.7&62.7 \\%
2g&1193&0.953&0.8326&--2&2.447&1.919&$\infty$&4.82&1.16&2.77&57.9 \\%
2h&1024&1.237&1.071&--2&2.888&2.877&$\infty$&4.13&1.5&3.4&54.5 \\%
2i&928.2&0.3380&0.8514&1.8&1.868&1&$\infty$&3.75&0.41&2.92&51.7 \\%
[1ex]%
4a&526.5&0.3&0.7555&1.757&2.108&1&$\infty$&3.59&0.364&4.43&50.4 \\%
4b&534.9&0.3&0.05205&1.556&2.386&5.239&$\infty$&4.7&0.364&6.26&57.8 \\%
4c&527&0.4147&1.293&1&2.233&1&$\infty$&4.09&0.503&5.21&52.9 \\%
4d&511&0.6507&0.1101&1&3&5.236&$\infty$&3.97&0.789&5.13&32.5 \\%
4e&584&0.3&0.1078&1.764&2.076&2.628&$\infty$&4.45&0.364&4.74&58.8 \\%
4f&491.3&0.3&0.3926&1.5&2.256&1.764&$\infty$&3.91&0.364&5.66&53.4 \\%
4g&549.6&0.3&0.7447&1.8&2.066&1&$\infty$&4.27&0.364&5.04&58.1 \\%
4h&510.6&1.237&3.313&--2&2.672&1.262&$\infty$&3.97&1.5&4.44&43.4 \\%
4i&423&0.3&2.799&1.282&2.336&1&$\infty$&3.29&0.364&3.95&30.5 \\%
 \hline
 \end{tabular}\par\medskip
Columns 2 to 8 give the values of the parameters obtained from the fitting
procedure; surface and volume densities are given in $\Msun\,\parc^{-2}$ and
$\Msun\,\parc^{-3}$ respectively. The last four columns give, in units of 
$10^{10}\Msun$, the total masses of disk and bulge, and the mass of the halo
within 10\kpc\ and 100\kpc.
\end{minipage}
\end{table*}

Tables \ref{tab-fit} and \ref{tab-best-fit} also describe variants of the
standard models. Models 2a and 2b explore the effect of changing $R_0$:
reducing $R_0$ from $8\kpc$ to $7.5\kpc$ increases $A$ and reduces $\vc(R_0)$.
The top two panels of Fig.~\ref{fig-vc-non-standard} show that it also changes
the form of $\vc(R)$ from steady decline at all $R>R_0$ to a steady rise at
$R\ga40\kpc$. This coupling between the value of $R_0$ and the rotation curve
at $R\gg R_0$ arises because the tangent velocities strongly constrain the mass
distribution at $R<R_0$. Hence any change in one component at $R<R_0$ must be
compensated by a change in another component. Since our model for the halo has
only a few free parameters, a change in its density at $R<R_0$ is accompanied
by significant changes in density at $R\gg R_0$, and vice versa. 

Model 2c shows the effect of fixing the inner slope of the halo's density
profile at $\gamma_\h=1$; this value is motivated by Navarro, Frenk \& White
(1996), who find that in simulations of cosmological clustering, dark-matter
halos tend to a universal profile that has a slope $\sim1$ at small radii.
Even though Model 2 has $\gamma_\h=-2$, fixing $\gamma_\h$ in this way has a
negligible effect on the observable properties of the model, and causes a
negligible increase in $\chi^2_{{}_\rmn{tot}}$. This is because the halo
makes a negligible contribution to the central density even for
$\gamma_\h=1$.  Moreover, the increase in $\gamma_\h$ is at small radii
largely compensated for by a six-fold increase in the halo's break radius
$r_{0,\h}$.

Model 2d explores the effect of fixing both the inner and the outer slopes
of the halo to values, $\gamma_\h=1,\beta_\h=3$, that are suggested by Navarro
\etal\ (1996). Again the model's observables and value of
$\chi^2_{{}_\rmn{tot}}$ change remarkably little, while $r_{0,\h}$ increases
again, this time by a further factor of $\sim3$.

Models 2e and 2f show the effect of employing a slightly non-exponential
disk: a term ${\pm}\,0.1\cos(R/R_\rmn{d})$ is added to the exponent in
equation (\ref{rho-disk}). The lower two panels of
Fig.~\ref{fig-vc-non-standard} show that at $R\la 10\kpc$ this extra term
changes the balance between the contributions to $\vc$ of the bulge, disk
and halo without significantly affecting the overall rotation curve. At
large $R$ the effect of the additional term is more dramatic in that with a
plus sign $\vc$ declines steadily at $R>R_0$, while with a minus sign it
rises gently at the largest radii. Again we encounter the consequence of the
rotation curve at large radii being inadequately constrained by the
observations: it mirrors changes in the functional form of the halo that are
designed to fit the data at $R\la R_0$.

Model 2g shows that increasing $\sigma_\rmn{BW}$, the target dispersion in
Baade's window, by $23\kms$ has little effects on the derived model: the
predicted dispersion is grudgingly raised from $108\kms$ to $115\kms$, and
$\chi^2_{{}_\rmn{tot}}$ increases significantly.

Microlensing surveys have suggested that the bulge may be significantly more
massive than studies of the tangent velocities would imply (e.g., Bissantz
\etal\ 1996). Model 2h shows the effect of imposing the constraint
$M_\rmn{bulge}\ge 1.5\times 10^{10}\Msun$. This causes $\vc$ to
be dominated by the bulge out to ${\sim}\,3.5\kpc$. It also, by the
mechanism described above, changes the form of $\vc(R)$ at
large $R$ so that with a more massive bulge $\vc$ is predicted to decline
significantly more steeply at $R\ga50\kpc$. Like the other model with an
artificially enhanced bulge (Model 2g), this model has an anamalously large
value of $\chi^2_{{}_\rmn{tot}}$.

\begin{table}
\def\mce{\multicolumn{2}{c}{-}}
\tabcolsep1mm
\caption[]{Parameter Correlations for models 2 (upper right) and 4 (lower left)}
	\label{tab-correl}
\begin{tabular}{l@{\hspace{3mm}}r@{.}lr@{.}lr@{.}lr@{.}lr@{.}lr@{.}lr@{.}l}
  & \multicolumn{2}{c}{$\rho_{0,\rmn{h}}$}
  & \multicolumn{2}{c}{$\gamma_\rmn{h}$}
  & \multicolumn{2}{c}{$\beta_\rmn{h}$}
  & \multicolumn{2}{c}{$r_{0,\rmn{h}}$}
  & \multicolumn{2}{c}{$r_{t,\rmn{h}}$}
  & \multicolumn{2}{c}{$\rho_{0,\rmn{b}}$}
  & \multicolumn{2}{c}{$\Sigma_\rmn{d,tot}$} \\ \hline
$\rho_{0,\rmn{h}}$  & \mce  &--0&999&--0&953&--0&999&--0&909&  0&917&--0&161 \\ 
$\gamma_\rmn{h}$    &--0&714& \mce  &  0&952&  0&998&  0&907&--0&920&  0&153 \\
$\beta_\rmn{h}$     &--0&990&  0&646& \mce  &  0&968&  0&989&--0&788&  0&112 \\
$r_{0,\rmn{h}}$     &--0&999&  0&705&  0&992& \mce  &  0&929&--0&900&  0&153 \\
$r_{t,\rmn{h}}$     &--0&963&  0&643&  0&984&  0&965& \mce  &--0&729&  0&109 \\
$\rho_{0,\rmn{b}}$  &--0&877&  0&312&  0&891&  0&882&  0&846& \mce  &--0&175 \\
$\Sigma_\rmn{d,tot}$&  0&089&--0&126&--0&061&--0&089&--0&005&--0&013& \mce   \\
\hline
\end{tabular}
\end{table}

It is not clear that a massive halo should be nearly round (axis ratio
$q_\h=0.8$) as in all the models described above, so Model 2i explores the
effect of assuming that the halo is highly flattened: axis ratio $q_\h=0.3$.
This has the effect of making the halo virtually a pure power-law component
that is dynamicaly dominant at all radii. Consequently, it predicts that the
rotation curve rises at the very largest radii. The massive, flattened halo
contributes strongly to $K_{z,1.1}$, so that the target value of $K_{z,1.1}$
is significantly overshot, despite the local disk surface density,
$\Sigma_0$, being pushed right down to its floor value,
$\Sigma_0=40\msun\parc^{-2}$. For this model $\chi^2_{{}_\rmn{tot}}$ is
on the high side.

Models 4a -- 4i explore the effect on Model 4 of the changes that were made
to Model 2 in making Models 2a -- 2i. Qualitatively the results are similar,
but quantitatively they tend to be smaller because the halo is very much
more important in Model 4 than it is in Model 2 and the effects of changes
in the disk and bulge are relatively minor.

Inspection of Table \ref{tab-best-fit} shows that there is a general
anticorrelation between the steepness of the halo's central density profile
and the bulge mass. In particular, when the halo density is strongly cusped
at the centre $(\gamma_\h>1$), the bulge mass is small: $M_{\rm
b}\sim0.37\times10^{10}\msun$.  Conversely, with the exception of the model
with the smallest value of $R_{\d,*}/R_0$, the bulge mass is large ($M_{\rm
b}\ga0.8\times10^{10}\msun$) when the halo has a hole at its centre
($\gamma_\h=-2$).

The asymptotic slope of the halo profile at large $R$ is always larger than
$\beta_\h=1.63$ and no model has a finite halo cut-off radius $r_{\rm t,h}$.
That is, the data imply only that the halo ends at $R>100\kpc$.

The only models that comes near to violating the lower limit
(\ref{local-sigma-nod}) on the local column density are  Model 1, which has
the smallest disc scale-length, Model 2h, which is obliged to have a massive
bulge, and Models 2i and 4i, which have highly flattened halos.

Table \ref{tab-correl} gives the correlation matrix of the fitted parameters
for both Models 2 and 4: the upper right triangle is for Model 2 while the
lower-left triangle is for Model 4. The parameters of the halo are strongly
correlated with one another. Indeed, the halo's density normalization
$\rho_{0,\rm h}$, central slope $\gamma_{\rm h}$ and scale length $r_{0,\rm
h}$ have almost unit correlations between them. The density normalization of
the bulge, $\rho_{0,\rm b}$ is strongly correlated with all the halo
parameters, especially $\rho_{0,\rm h}$, $\gamma_{\rm h}$ and $r_{0,\rm h}$.
By contrast, $\Sigma_{\rm d,tot}$ is only weakly correlated with the other
parameters.  These strong correlations between parameters reflect the
existence of a long, level valley in parameter space.

\section{CONCLUSIONS}\label{sec-conclude}
We have fitted a multiparameter mass model to the available kinematic data
for the Milky Way. The wide variety of models that emerge from this fitting
process demonstrates that the mass distribution within the Milky Way is
currently ill determined.

The principal problem in determining the Galaxy's mass distribution is that
very few hard facts are available regarding the vertical distribution of the
Galaxy's mass. Also the circular speed is observationally much less well
constrained at $R>R_0$ than it is at $R<R_0$. These deficiencies
in the spatial coverage of the data have several unfortunate effects. First
they oblige use to represent the Galaxy as a superposition of components,
and to adopt simple functional forms for the distribution of mass within
each component. 

The assumed distribution of mass within the halo plays a particularly
important and confusing role. Since we know nothing about the halo except
what can be gleaned from studies such as this, we have allowed the halo
density as much freedom as is compatible with (i) its being much thicker
than the disk (axis ratio $q_\rmn{h}\ge=0.3$), (ii) its density function
containing only a few free parameters, and (iii) its not being implausibly
sharp-edged ($-2\le\gamma_\rmn{h}\le1$, $\beta_\rmn{h}\ge1$).

With these assumptions we find that even the circular-speed curve of the
best-fitting model depends significantly on both the adopted values of
parameters such as $R_{\rm d,*}/R_0$ and $R_0$, and on the adopted
functional forms of the components. The density model is even less well
constrained by the data.  In particular, remarkably small changes in
$\vc(R)$ and the other observational constraints can be associated with
dramatic changes in the distribution of mass between the different
components, and the degree of central concentration of the halo.

A particularly disturbing phenomenon is that a change in a component that
contributes to $\vc$ only at small radii causes the predicted value of $\vc$
to change at large radii. This connection between small and large scales
within the Galaxy is established by the halo component, which must be
allowed to contribute to the density at all radii and yet be determined by a
small number of parameters. In general there is a clear need to model
components non-parametrically or at least by functions that contain more
parameters. However, we do not yet have enough observational constraints to
constrain adequately models that are significantly more complex than those
used here.

As is traditional, we have represented the Galaxy as a superposition of
components that individually represent plausible stellar systems. The
justification for the use of such components in preference to a family of
orthogonal (and therefore non-positive) functions, is the hypothesis that
these components do, in fact, represent real physical systems. While there
can be no doubt of the reality of the disk and bulge, the status of the halo
is entirely speculative. Indeed, although observations of external galaxies
clearly require high mass-to-light ratios at large radii, it does not follow
that these reflect the existence of a physically distinct dark halo; it is
perfectly possible that the mass-to-light ratio of the disk or bulge
increases strongly away from the Galactic centre.  If the halo does
represent a distinct dynamical entity, then in an exercise like the present
one it should emerge with a dynamically plausible density profile. Table
\ref{tab-best-fit} shows that it frequently fails this test: in just over
half the models, the central density slope $\gamma_{\rm h}$ lies at one or
other extreme of its permitted range $-2\le\gamma_{\rm h}\le1.8$. In four
models (Models 3, 2a,f,i) the halo's density profile is an essentially
featureless power law of slope $\sim-1.8$. In seven models the halo has a
hole at its centre. $N$-body simulations of structure formation offer no
encouragement to the idea that the Galaxy's distribution of axions or other
exotic particles would have a hole at its centre. Nor do they suggest that
it should be a featureless power law (Navarro \etal. 1996).

The road to an improved understanding of the Galaxy's mass distribution must
lie with the introduction of more observational constraints, especially ones
that relate to $R>R_0$ and $z\ga R$. The goal of later papers in
this series is to bring to bear on this problem observations of halo stars,
which should provide a wealth of information about the density at
$z\ga R$, and, to a lesser extent, about the density at $R>R_0$.
The models described here simultaneously provide starting points for this
enterprise and demonstrate its urgency by underlining that at the
present time we know depressingly little about the distribution of mass
within our own Galaxy.

%REFERENCES

\end{document}